\documentclass[aps,prL,twocolumn,floats]{revtex4}
\usepackage[margin=0.8in,footskip=0.25in,paperwidth=8in,paperheight=13.5in]{geometry}
\usepackage{latexsym}
\usepackage{amsmath}

\usepackage{amssymb}
\usepackage{bm}
\usepackage{wasysym}
\usepackage[dvips]{color}
\usepackage{graphicx}
\usepackage{subfigure}
\usepackage{epsfig}
\usepackage{pgfplots}
\usepackage{balance}
\DeclareMathAlphabet{\mathpzc}{OT1}{pzc}{m}{it}
\newcommand{\hide}[1]{}
\newcommand{\veps}{\varepsilon}

\def\ra{\rangle}
\def\la{\langle}

\def\veps{\varepsilon}

\usepackage{tikz}
\usetikzlibrary{matrix}
\usetikzlibrary{arrows}
\graphicspath{{./figures/}}
\usepackage{float}
\usepackage[colorlinks,bookmarks=true,citecolor=blue,linkcolor=blue,urlcolor=blue, breaklinks=true]{hyperref}
\usepackage[sort&compress]{natbib}
\usepackage[hyphenbreaks]{breakurl}
\begin{document}
\title{Klein Bound States in Single-Layer Graphene}
\author{Y. Avishai$^{1,3,4}$ and Y. B. Band$^{1,2}$}
\affiliation{$^1$Department of Physics and The Ilse Katz Center for Nano-Science, Ben-Gurion University of the Negev, Beer-Sheva, Israel.\\
  $^2$Department of Chemistry, Ben-Gurion University of the Negev, Beer-Sheva, Israel.\\
  $^3$New-York Shanghai University, 1555 Century Avenue, Shanghai, China. \\
  $^4$Yukawa Institute for Theoretical Physics, Kyoto, Japan.\\
}

\begin{abstract}
The Klein paradox, first introduced in relation to chiral tunneling, is also manifested in the study of bound-states in single-layer graphene with a 1D square-well potential. We derive analytic (and numerical) solutions for  bound-state wavefunctions, in the absence and in the presence of an external transverse magnetic field, and calculate the corresponding dipole transition rates, which can be probed by photon absorption experiments.  The role of parity and time-reversal symmetries is briefly discussed. Our results are also relevant for the physics of bound states of light in periodic optical waveguide structures.
\end{abstract}

\date{\today}

%\pacs{72.80.Vp, 71.15.-m, {\color{red}add?}}

\maketitle
 
{\it Introduction}.---
Chiral tunneling of electrons through a 1D potential barrier in single layer graphene was first considered in a seminal paper by Katsnelson, Novoselov, and Geim \cite{Katsnelson_06}. A closely related and physically motivated problem concerns the formation of electron bound states in a 1D (symmetric) potential well \cite{Peeters, Barbier, Ramezani, Nguyen_09, Gutierrez_16} (bound-states here refers to bound in one direction and free in the other direction). In this Letter we elucidate several novel aspects of such bound-states amenable to experimental verification.  Our main results are:  (1) In the absence of a magnetic field, bound-state eigenfunctions and eigenvalues are derived analytically, and  electric dipole transition strengths are calculated to determine the absorption spectrum between bound-states. Parity and time reversal symmetry are employed to find the relation between the two (pseudo)-spinor components. (2) In the presence of an external magnetic field, analytic expressions for the bound state wavefunctions for a discrete sequence of potential strengths are derived, and are used to determine the measurable areal densities and currents.  Based on ideas presented in Refs.~\cite{Longhi_10, Longhi_11}, our formalism also applies to the occurrence of bound states of light in periodic optical waveguide structures.

\begin{figure}%[htb]
\subfigure[]
{\includegraphics[width=7.5cm,height=6.0cm]{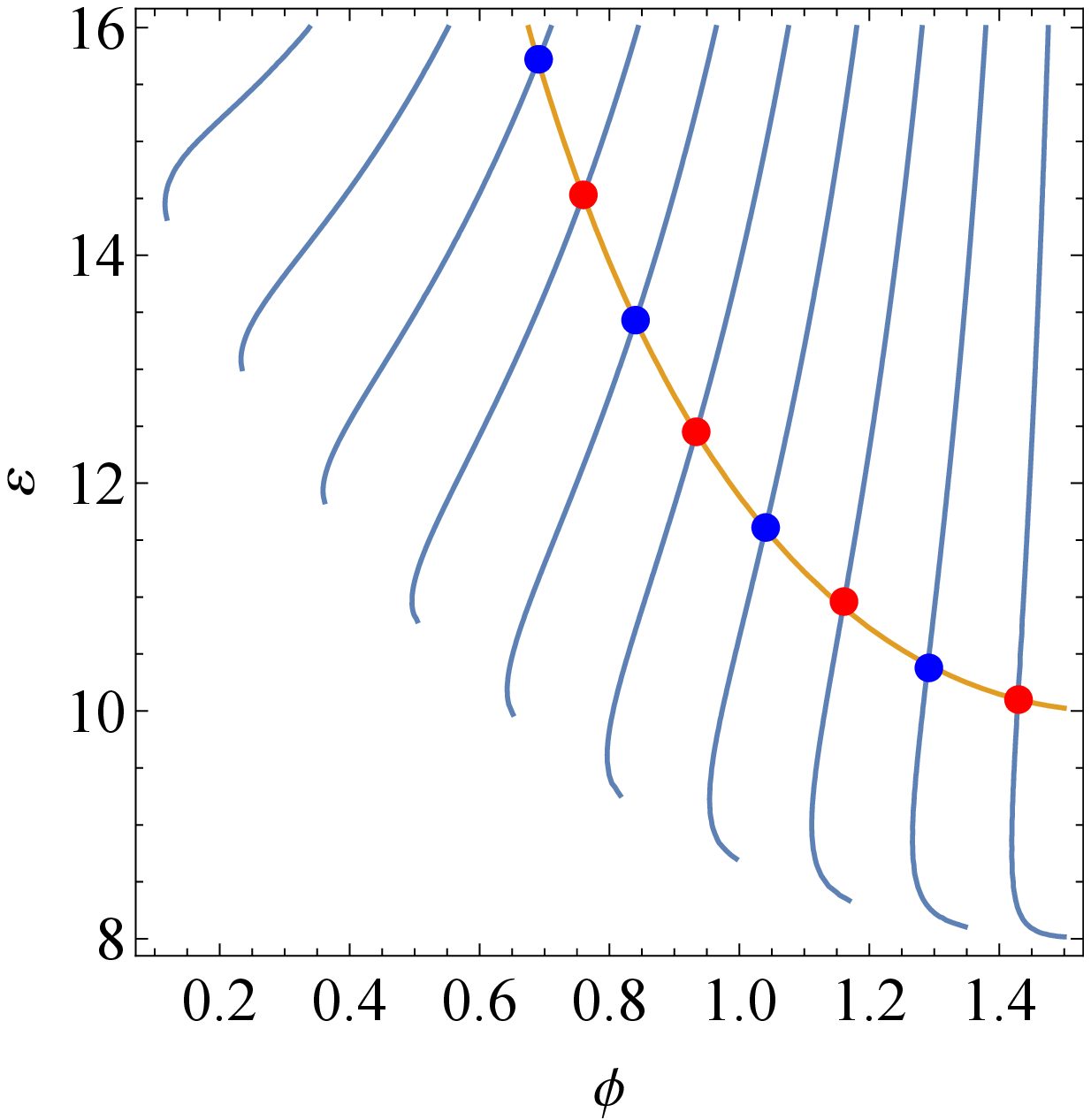}}
%{\includegraphics[height=0.25\linewidth,angle=0] {Square-Well-Fig1a.eps}}
\ \ \ \subfigure[]
{\includegraphics[width=7.5cm,height=5.25cm,angle=0] {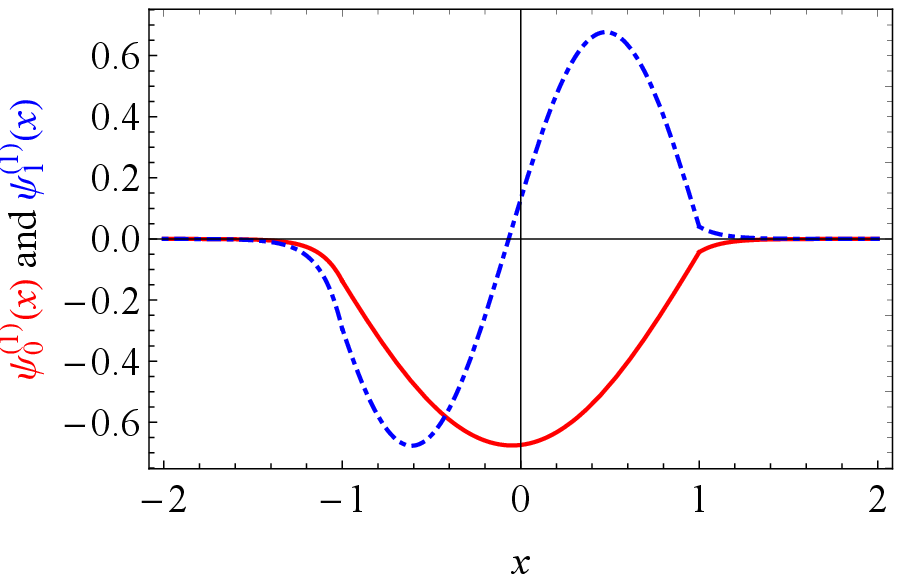}}
\caption{(a) Nodes of det$[A(\veps, \phi)]$, Eq.~(\ref{MatrixA}), in the $(\phi,\veps)$ plane (blue curves), and the curve $k_y=\veps \sin \phi=10$ (orange curve).  The pairs $(\veps_n,\phi_n)$
specified by blue and red dots are the bound state energies for $k_y=\veps_n\sin \phi_n=10.$   
 (b) Upper components of $\psi_0(x)$ (red solid curve) and $\psi_1(x)$ (blue dot-dashed curve) versus $x$. The lower components are related to the upper ones via Eq.~(\ref{gswf}). Note that the wavefunctions do not have a definite symmetry around $x=0$ (see discussion on the role of parity below). }
\label{bscontourplot}
\end{figure}

%{\color{red} Remove due to lack of space.  IS IT TRUE WE DON'T HAVE SPACE??? 
%---------------------------------------------}\\
%Graphene can be treated as two interpenetrating triangular lattices often labeled by $A$ and $B$. In the presence of an external potential $U(x,y)$, dynamics of the low-energy quasi-particles of the system near the Dirac points is governed by the 2D Dirac Hamiltonian for massless particles: $H = v_F {\boldsymbol \sigma} \cdot {\bf p} + U(x,y)$ \cite{Katsnelson_06}.  Here ${\boldsymbol \sigma}$ is the pseudospin Pauli matrix vector, ${\bf p} = (p_x, p_y)$ is the 2D momentum operator and ${\bf v}_F$ is the Fermi velocity.\\
%{\color {red} End remove--------------------------------------}\\
{\it Bound states in a symmetric 1D square-well}.---
We search for bound states of a massless particle in single-layer graphene using the 2D Dirac equation with 1D square-well symmetric potential $U(x)=U_0 \Theta(\vert x \vert -L)$. Employing $L$ as a length unit, we define dimensionless coordinates $x \to x/L, \ \ y \to y/L$, potential $u(x)=L U(x)/(\hbar v_F) \equiv u_0 \Theta(\vert x \vert -1)$, energy $\veps=L E/(\hbar v_F)$, (where $E$ is the energy in physical units), and wavenumber $k=\veps$, (where $E/(\hbar v_F)$ is the Fermi wavenumber in physical units).  Klein physics \cite{Klein} occurs for $u_0> \veps >0$ where inside the well ($\vert x \vert <1$) the Fermi energy lies in the conduction band while outside the well ($\vert x \vert > 1$) the Fermi energy lies in the valence band \cite{Klein, AF_11}. 
Near the ${\bf K}'$ Dirac point, the time-independent 2D Dirac equation (in dimensionless variables) is,
\begin{equation} \label{1y}
{\cal H}\Psi \equiv [-i(\sigma_x \partial_x+\sigma_y \partial_y)+u(x)]\Psi(x,y) = \veps  \Psi(x,y).
\end{equation} 
\vspace{-0.0in}
Under parity transformation $(x,y) \to (-x,y)$, the potential is symmetric, $u(x)=u(-x)$, but the total Hamiltonian is not, ${\cal H}(-x,y) \ne {\cal H}(x,y)$. 
The general solution of the wavefunction in the three different regions is, $ \Psi(x,y)=e^{i k_y y} \psi(x)$,
\begin{equation} \label{psi-bound-state-x}
\psi(x)\mbox{=}\begin{cases} a \binom{1}{e^{i \phi}} e^{i k_x x}+b \binom{1}{-e^{-i \phi}} e^{-i k_x x} \ \ (|x|<1), \\
\alpha \binom{1}{-e^{i \theta}} e^{i q_x x}+\beta \binom{1}{e^{-i \theta}} e^{-i q_x x} \ \ (x>1), \\ 
\gamma \binom{1}{-e^{i \theta}} e^{i q_x x}+\delta \binom{1}{e^{-i \theta}} e^{-i q_x x} \ \ (x<-1),
\end{cases}
\end{equation}
where $\phi$ is the inclination angle and $\theta$ is the refractive angle.  The dimensionless momentum vector inside the well [where $u(x)=0$] is
\begin{equation} \label{kF}
{\bf k}=\veps(\cos \phi ~ \hat{\bf x}+\sin \phi ~ \hat {\bf y})\equiv k_x \hat{\bf x}+k_y  \hat {\bf y},
\end{equation}
and $\vert {\bf k} \vert=\veps=\sqrt{k_x^2+k_y^2}$.  The $x$ component of the momentum outside the well [where $u(x) = u_0 > 0$] and the refractive angle are given by
\begin{eqnarray} \label{qx}
&&  q_x=\sqrt{(\veps-u_0)^2-k_y^2}, \nonumber \\
&& \tan \theta=\frac{k_y}{q_x}=\frac{\veps \sin \phi}{\sqrt{(\veps-u_0)^2-(\veps \sin \phi)^2}}.
\end{eqnarray}
In the $p$-$n$-$p$ junction analyzed here, Klein tunneling occurs for $u_0/(1+\sin \phi)>\veps >0$ (where $q_x$ is real), whereas Klein bound states occur for $u_0>\veps > u_0/(1+\sin \phi)>0$, for which
\begin{equation}  \label{qim}
  q_x=i \kappa_x(\veps,\phi)\equiv i \sqrt{(\veps \sin \phi)^2 - (u_0-\veps)^2},
\end{equation}
and $\kappa_x(\veps,\phi)>0$.  The bound state wavefunctions must decay exponentially as $e^{-\kappa_x \vert x \vert}$ as  $\vert x \vert \to \infty$. In this region $\tan \theta= -i k_y/\kappa_x$ is pure imaginary, and $\tan^2 \theta <-1$. Consequently, $\sin \theta$ is real and $\cos \theta$ is imaginary. 
To insure asymptotic decay at large $\vert x \vert$ we must set $\beta=\gamma=0$ in Eq.~(\ref{psi-bound-state-x}). Continuity of $\psi(x)$ at $x=\pm 1$ yields a homogeneous  system of four linear equations for the complex coefficient vector ${\bf c} \equiv (a, b, \alpha, \delta)^T$ that is an eigenvector with zero eigenvalue of the matrix $A(\veps)$, $A(\veps){\bf c}=0$ ($A(\veps)$ is explicitly given in the Supplemental Material (SM) \cite{SM}).  The determinant of $A(\veps)$ is given by
\begin{eqnarray} \label{MatrixA}
&& C \, \mbox{det}[A(\veps)] = \kappa_x(\veps,\phi) \cos \phi \cos(2 \veps \cos \phi) \nonumber \\
&&+[\veps(1+\sin^2 \phi)-u_0]\sin(2 \veps \cos \phi),
\end{eqnarray}
where $C$ is a non-vanishing multiplicative constant and the expression on the RHS is real.
Bound-states occur at energies $\veps_n$ for which $\mbox{det}[A(\veps_n)] = 0$.  
For reasons that will be explained later, we focus on bound states at different energies $\{ \veps_n \}$ but for {\it the same} $k_y=\veps_n \sin \phi_n$.  We use the following realistic values for the parameters of graphene: $L=172$ nm, $U_0 = 50$ meV, which yields $u_0=\frac{LU_0}{\hbar v_F} \approx 16.0$.
The pattern of bound state energies in the $(\phi,\veps)$ plane is shown in Fig.~\ref{bscontourplot}(a), together with the curve $k_y=\veps \sin \phi=10$ (= 0.0581 nm$^{-1}$). The intersection points indicate bound-state energies $\{ \veps_n \}$ with the same value of $k_y=\veps_n \sin \phi_n$. 
%{\color{red} Mathematica file: Klein-Modified-Determinant-band2.nb} 
%Klein-Bound-State-Matrix-Eigensystem4-ky-5

{\it Bound state wavefunctions}.---
Now we compute the wavefunctions for  $(\veps_n,\phi_n)$, $n=0, 1, \ldots, 7$, see Fig.~\ref{bscontourplot}(a).  The pairs $(\veps_n,\phi_n)$ are inserted into the matrix $A$ and the spinor bound-state wavefunctions $\psi_n(x)=\binom{\psi_n^{(1)}}{\psi_n^{(2)}}$ are determined in terms of the four coefficients ${\bf c}_n \equiv (a_n,b_n,\alpha_n,\delta_n)$, i.e., the solution of the eigenvalue equation $A(\veps_n,\phi_n){\bf c}_n^T=0$.  
Due to parity symmetry (see below), the components of the spinors are subject to the constraints,
%{\color{red} Mathematica file: Proof-Of-Symmetry1.nb}
\begin{eqnarray} \label{wfsymmetry}
&& \mbox{Im}[\psi_n^{(1)}(x)]=\mbox{Re}[\psi_n^{(2)}(x)]=0, \nonumber \\
&& \mbox{Im}[\psi_n^{(2)}(x)]=(-1)^n\mbox{Re}[\psi_n^{(1)}(-x)].
\end{eqnarray}  
Analytic expressions for the ground and excited state wavefunctions are derived by choosing
\begin{equation} \label{abground-excited}
a=b^*={\cal A}_ne^{i \eta_n}, \ \ \eta_n= (2n+1)\tfrac{\pi}{4}-\tfrac{1}{2}\phi.
\end{equation}
where ${\cal A}_n$ are real normalization constants and the phase $\eta_n$ is chosen to satisfy the symmetries in Eq.~(\ref{wfsymmetry}). Combining Eqs.~(\ref{psi-bound-state-x}) and (\ref{abground-excited}), the bound-state wavefunctions, $\Psi_n(x,y)=e^{i k_y y}\psi_n(x)$  for $\vert x \vert < 1$ are, 
%{\color{red} Mathematica file: Proof-Of-Symmetry.nb}
%{\color {red} Klein-Bound-State-Matrix-Eigensystem4-ky-5}
\begin{eqnarray} \label{gswf}
&& \psi_n(x,y)=  \mathcal{A}_n \binom{\psi_n^{(1)}(x)}{(-1)^n i\psi_n^{(1)}(-x)} \nonumber \\
&&=\mathcal{A}_n\binom {\cos[\gamma_n^-(x)]
+ \sin[\gamma_n^-(x)]}
{ (-1)^n i \{\cos[\gamma_n^+(x)]
+ \sin[\gamma_n^+(x)]\}} , \\ \nonumber
&& \gamma_n^\pm(x)=\tfrac{1}{2}(\phi_n \pm 2 k_{nx} x),
%&&  \Psi_1(x,y) \! = \! \mathcal{A}_1e^{i k_y y}\binom{ -\cos[\tfrac{1}{2}(\phi-2 (\veps_1 \cos \phi) x)]
%+ \sin[\tfrac{1}{2}(\phi-2 (\veps_1 \cos \phi) x)]}
%{i \{\cos[\tfrac{1}{2}(\phi+2 (\veps_1 \cos \phi) x)]
%-\sin[\tfrac{1}{2}(\phi+2 (\veps_1 \cos \phi) x)]\}}= \mathcal{A}_1e^{i k_y y} \binom{\psi_1^{(1)}(x)}{-i\psi_1^{(1)}(-x)}.
\end{eqnarray}
where $k_{nx}= \veps_n \cos \phi_n$. The decaying parts of the wavefunctions for $\vert x \vert >1$ are determined by the coefficients $\beta,\delta$, and the symmetry specified in Eq.~(\ref{wfsymmetry}) is fulfilled {\it for all} $x$. The two upper components of the spinor wavefunctions $\psi_{n=0,1}(x)$  are shown in Fig.~\ref{bscontourplot}(b).

The symmetry specified in Eq.~(\ref{wfsymmetry}) also implies that
$\psi_0^\dagger(x) \psi_1(x)$
is an odd function of $x$.  Hence, $\langle \psi_0 \vert \psi_1 \rangle=0$, i.e., the two states are orthogonal, as are any two different eigenfunctions.
 
 {\it Currents}.---
Bound states, with wavefunctions $\psi_n(x)=\binom{\psi_n^{(1)}(x)}{\pm i\psi_n^{(1)}(-x) }$, do not carry current along $x$:
$J_{nx}(x) \equiv \psi_n^\dagger(x) \sigma_x \psi_n(x)=0$.
However, they do carry current along $y$, $J_{ny}(x)\equiv \psi_n^\dagger(x) \sigma_y \psi_n(x) \ne 0$, $(n=0,1)$,
that is symmetric under $x \leftrightarrow -x$ and it quickly decays for $\vert x \vert > 1$. As we discuss below in connection with time reversal invariance, all states are Kramers degenerate, and the two degenerate states forming a Kramers pair carry currents in opposite directions.

{\it Parity}.---
The importance of parity in the physics of graphene is discussed in %, e.g., 
Ref.~\cite{Riazuddin_18}, where it is shown that parity operator in (1+2) dimensions plays an interesting role and can be used for defining conserved chiral currents (see also Ref.~\cite{Sadurni_15}).  Here we concentrate on bound states, wherein the current along $x$ should vanish, and consider the role of the parity transformation under which the Hamiltonian is {\em not} invariant.
For a symmetric potential, $u(x)=u(-x)$, we consider the static (time-independent) case with Hamiltonian ${\cal H}(x,y)$ introduced in Eq.~(\ref{1y}). 
The parity transformation in 2+1 dimensions is taken to mean the transformation $(x,y) \to (-x,y)$.  For massless Dirac fermions this transformation is realized by the operator $\sigma_y$.  Explicitly, 
\begin{eqnarray} \label{HP}
&& {\cal H}^P(x,y) \equiv \sigma_y {\cal H}(x,y) \sigma_y
=i \sigma_x \partial_x-i \sigma_y \partial_y+u(x) \nonumber \\
&&= {\cal H}(-x,y) \ne {\cal H}(x,y).
\end{eqnarray} 
Thus, near a given Dirac point, say ${\bf K}'$, ${\cal H}$ is not parity invariant [despite the fact that $u(x)=u(-x)]$ \cite{footnote1}.  
However, for a symmetric potential the wavefunctions $\psi_n(x)$ in Eq.~(\ref{gswf}) obey the symmetry relations,
\begin{equation} \label{sigma-psi}
\sigma_y \psi_0(x)=\psi_0(-x), \quad \sigma_y \psi_1(x)=-\psi_1(-x).
\end{equation}
Equation (\ref{sigma-psi}) is a concrete realization of Eq.~(14) in Ref.~\cite{Riazuddin_18}. 
Hence, we define $\psi_n(x)$ as being $\binom {\text{even}} {\text{odd}}$ under parity if and only if $\sigma_y \psi(x)=\pm \psi(-x)$. With this assignment, Eq.~(\ref{sigma-psi}) is consistent with (albeit different than) the non-relativistic one-dimensional problem, where, in a symmetric potential, the parity of eigenstates is such that $\psi_n(-x)=(-1)^n \psi_n(x), \ n=0,1,2,...$, and the ground-state is symmetric. By definition, 
\begin{equation} \label{HPpsi}
{\cal H} \psi_n(x)=\veps_n \psi_n(x) \Rightarrow \ {\cal H}^P \psi_n(-x)=\veps_n \psi_n(-x). 
\end{equation}  
Thus, $\psi_n(x)$ and $\psi_n(-x) \ne \pm \psi_n(x)$ are respectively eigenfunctions  of ${\cal H}$ and
 ${\cal H}^P \ne {\cal H}$ with the same eigenvalue $\veps_n$.

{\it Time Reversal Invariance}.---
The time reversal operator is ${\cal T}=i \sigma_y K$, where $K$ is the complex conjugation operator.  It is easy to check that $[{\cal H},{\cal T}]=0$, so that each state is doubly (Kramers) degenerate.  Applying the operator ${\cal T}$ on a wavefunctions $\Psi_n(x,y)$, Eq.~(\ref{gswf}) we obtain [recall that $\psi_n^{(1)}(x)$ is real and $\psi_n^{(2)}(x)=(-1)^{n}i\psi_n^{(1)}(-x)$ is purely imaginary], 
\begin{equation} \label{Tpsi} 
\Psi_n^{\cal T}(x,y)=
%{\cal T}  {\cal A}_ne^{i k_y y}  \binom{\psi_n^{(1)}(x)}{(-1)^{n}i \psi_n^{(1)}(-x)} =
 {\cal A}_ne^{-i k_y y} \binom{(-1)^n\psi_n^{(1)}(-x)}{i\psi_n^{(1)}(x)}, 
\end{equation}
which is the Kramers partner of $\Psi_n(x,y)$, i.e.,
%\begin{equation} \label{Kramers}
%\end{equation}
${\cal H} \Psi_n^{\cal T}(x,y)=\veps_n \Psi_n^{\cal T}(x,y)$.

\begin{figure}%[htb]
\includegraphics[width=0.9\linewidth,angle=0] {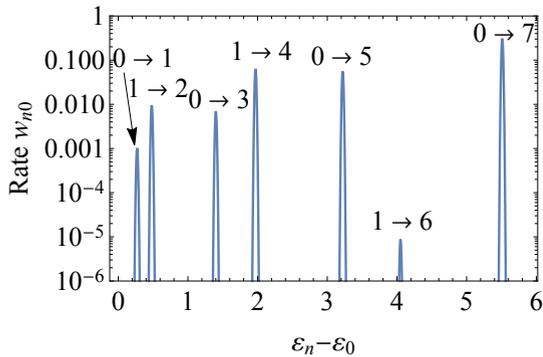}
\caption{Absorption spectrum of the transitions $0 \to 1$, $0 \to 3$, $0 \to 5$, $0 \to 7$, $1 \to 2$, $1 \to 4$, $1 \to 6$. The absorption rate $w_{mn}$ (in dimensionless units) from level $n$ to level $m$ is plotted as a function of the resonant absorption photon energy $\hbar \omega_{nm} = \veps_n -\veps_m$ (in dimensionless units).}
\label{absorption_spectrum}
\end{figure}

{\it Electromagnetic Transitions}.---
%{\color{red} Klein-Bound-State-Matrix-Eigensystem4-ky-5.nb}
Consider $E1$ transitions induced by $x$ polarized light such that the dipole operator is ${\cal O}(x)= e E_x x$, where $E_x$ is the electric field amplitude. The parity of the product $\psi_n^\dagger(x) \psi_m(x)$ is $(-1)^{n+m+1}$.  Because $k_y$ is conserved and is the same for $\Psi_{n}(x,y)$ and $\Psi_{m}(x,y)$, we have,
\begin{equation} \label{ME}
 \la \Psi_{m}\vert {\cal O} \vert \Psi_{n} \ra = \tfrac{1}{2}[1-(-1)^{n+m}] eE_x\la x \ra_{n,m}. 
\end{equation} 
Figure~\ref{absorption_spectrum} shows the absorption spectrum of the transitions  $0 \to 1$, $1 \to 2$, $0 \to 3$, $1 \to 4$, $0 \to 5$, $1 \to 6$, $0 \to 7$, where the absorption rates  (in arbitrary units) $w_{nm}$ from $m$ to $n$ are proportional to $\omega_{nm}^4 \, \left| \langle \psi_n| x |\psi_m \rangle \right|^2$ where $\hbar \omega_{nm} = \veps_n -\veps_m$ \cite{Band_Avishai}. 

Strictly speaking, electrons can occupy orbits with arbitrary $k_y<\veps$ and transitions can occur between the pertinent energy states. However, practically, an experiment can be carried out in a graphene nano-ribbon of width $L_y$ such that $k_y=\frac{2 \pi p}{L_y}, (p=1,2,\ldots)$ is quantized.  If $L_y$ is small enough, only the lowest mode is occupied. In our example, $k_y=10$ and $\veps<u_0 = 16$ (in dimensionless units). If this value of $k_y$ corresponds to the lowest mode $p=1$, then the second mode ($p=2$) has $k_y=20>\veps$.  In physical units, this implies $k_y=0.058$ nm$^{-1}$ and $L_y=108$ nm. Experimental fabrications of much lower nano-ribbon width have already been reported \cite{Barone}.
%{\color{red} (Klein-Bound-State-Matrix-Eigensystem4-ky.nb)}
%{\color{red} Mathematica file: Proof-Of-Symmetry.nb}

{\it Bound States in a perpendicular magnetic field and square well}.---
Analysis of bound states in the presence of uniform perpendicular magnetic field and a square well potential enables an access to ``un-quantized'' Landau functions in graphene.  First recall the extensively studied case $U(x)=0$ (see {\it{e.g.}}, Ref.~\cite{Zala}).  
In the Landau gauge, $A_y = Bx$, the spinor wavefunction is $\Psi(x,y) = e^{i k_y y} \psi(x)$.  Introducing the magnetic length $\ell=\sqrt{\hbar c/(e B)}$ enables formulation in terms of the dimensionless position, wave number and binding energy: $x \to x/\ell$,  $k_{x,y} \to k_{x,y} \ell $ and $\veps=\frac{\ell E}{\hbar v_F}$. 
The bare equation with dimensionless variables and parameters then reads, $[-i \sigma_x \partial_x + \sigma_y(k_y- x)]\psi(x) = \veps \psi(x)$.  It is simplified after a shift and scaling of the position coordinate, $x \to \frac{z}{\sqrt{2}}+k_y$,
%{\color{red} Klein-x-z-transformation.nb \, \, \, Klein-Magneic-Check-Equation-z.nb}
\begin{equation} \label{mag8a}
 {\cal H} \psi(z)\equiv[-i \sigma_x \partial_z -\tfrac{1}{2}z\sigma_y ]\psi(z)=\veps \psi(z),
\end{equation} 
whose general solution is (with $\bar{\delta} \equiv 1-\delta$), 
\begin{equation} \label{mag8}
  \binom{\psi^{(1)}(z)} {\psi^{(2)}(z)}\!=c_1 \binom{D_{\nu_1}(z)}{ \tfrac{\veps}{i}D_{\nu_1-1}(z)}
+c_2 \bar{\delta}_{\veps,0}\binom {D_{\nu_2}(iz)} {\tfrac{-1}{\veps}D_{\nu_2+1}(i z)} 
\end{equation}
where $D_{\nu}(z)$ is the parabolic cylinder function, 
$z \equiv z(x)=\sqrt{2}(x-k_y)$, $\nu_1=\veps^2$, $\nu_2=-(\veps^2+1)$.  If the wavefunction is required to be square integrable on the whole interval $-\infty < z < \infty$, we must set $\veps^2=n$, (where $n$ is a non-negative integer), and $c_2 = 0$ (because wavefunctions with imaginary arguments blow up). These constraints determine the Landau quantized energies $\veps=\pm \sqrt{n}$ and wavefunctions for electrons in graphene.

In the scaled shifted variable $z$ the square-well potential 
$U(x)=U_0 \Theta (\vert x \vert-L)$ reads, 
\begin{equation} \label{mag12} 
u(z) = \begin{cases} 0,  \ \ z(-L) < z < z(L) \\ u_0, \ \ \ \ \ \mbox{otherwise} \end{cases}, 
\end{equation}
where $u_0=\frac{\ell U_0}{\hbar v_F}$ and $z(L)=\sqrt{2}L-k_y \equiv L_1$,
$z(-L)=-\sqrt{2}L-k_y \equiv L_2 \ne -z(L)=-L_1$, hence $k_y=-\frac{1}{2} [z(L)+z(-L)]$. Thus, a symmetric well in $x$ is not symmetric in $z$.  The eigenvalue problem is specified by the 
set of equations defined for $-\infty < z < \infty$,
\begin{equation} \label{mag8a}
[-i \sigma_x \frac{d}{d z} -\tfrac{1}{2}z\sigma_y ]\psi(z) = \begin{cases} \veps \psi(z),
 \ \ z \in [L_2 , L_1] \\ (\veps-u_0) \psi(z), \ \ z \notin [L_2 , L_1]. \end{cases} 
\end{equation} 
Here $\psi(z)=\binom{\psi^{(1)}(z)}{\psi^{(2)}(z)}$, and $\veps$ is the energy eigenvalue that needs to be determined.  As in Eq.~(\ref{mag8}), the solutions can be expressed in terms of parabolic cylinder functions $D_\nu(\cdot)$, and the spinor wavefunction is required to be continuous everywhere and square-integrable.  For $z \in [L_2,L_1]$ the solution reads,
\begin{equation} \label{center}
  \psi_{\text{c}}(z)=c_1 \begin{pmatrix} \! D_{\nu_1}(z) \\ 
  -i \veps D_{\nu_1-1}(z) \end{pmatrix}
  +c_2 \bar{\delta}_{\veps,0}\begin{pmatrix}\! D_{\nu_2}(i z)\\-\frac{1}{\veps}D_{\nu_2+1}(i z) 
  \end{pmatrix}
\end{equation}
Generically, the orders
$\nu_1=\veps^2$, $\nu_2=-(\veps^2+1)$ in Eq.~(\ref{center}) are not (non-negative) integers.
In the external regions $z \notin [L_2,L_1]$, the only solutions of the second of Eq.~(\ref{mag8a}) that decay as $\vert z \vert \to \infty$ are such that: (1) the order $\nu$ of $D_\nu(\cdot)$ should be a non-negative integer, and (2) the argument of $D_\nu(\cdot)$ must be real. The most general solution is then an infinite linear combination of Landau functions $L_{sn}(z)=\binom{D_n(z)}{s i \sqrt{n}D_{n-1}(z)}, \ n=0,1,\ldots, \ s=\mp$.  A
general numerical solution is worked out in the supplemental material\cite{SM}. Here we show that analytic solutions exist for specific discrete values of the potential strength $u_0$.  We employ the following solutions of Eq.~(\ref{mag8a}) for $z \notin [L_2,L_1]$, 
with $\veps = u_0 \pm \sqrt{n}$, that is,  $n=(\veps-u_0)^2$:
\begin{eqnarray} \label{right-left}
&&\psi_{\text{right}}(z)= c_{3} \Theta (z-L_1) \begin{pmatrix} \! D_{n}(z), \  \\ 
  \pm i \sqrt{n} D_{n-1}(z) \end{pmatrix}, \nonumber \\
  &&\psi_{\text{left}}(z)= c_{4} \Theta(L_2-z) \begin{pmatrix} \! D_{n}(z) \\  \pm i \sqrt{n} D_{n-1}(z) \end{pmatrix}.
  \end{eqnarray}

\begin{figure}[htb] 
\subfigure[]
{\includegraphics[width=0.35 \textwidth,angle=0] {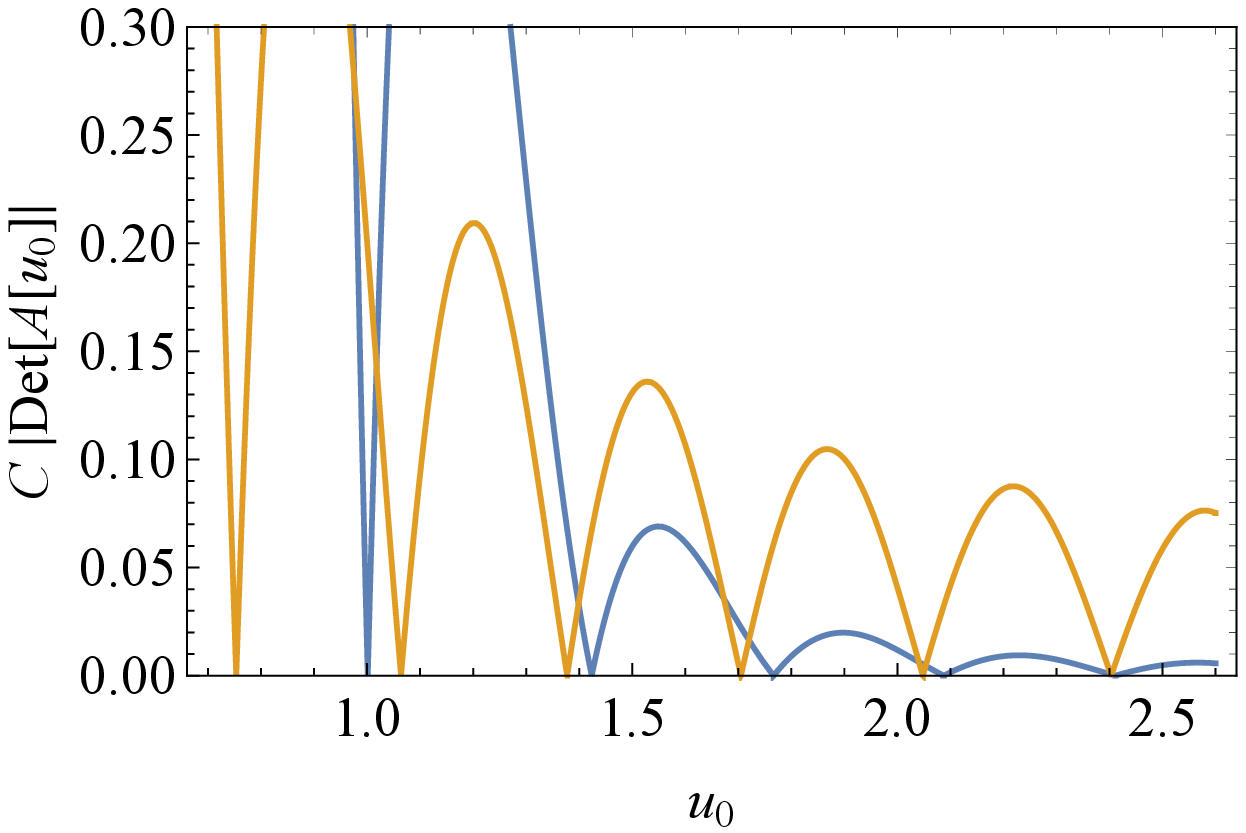}}
\ \ \ \ \ \ \subfigure[]
{\includegraphics[width=0.35 \textwidth,angle=0] {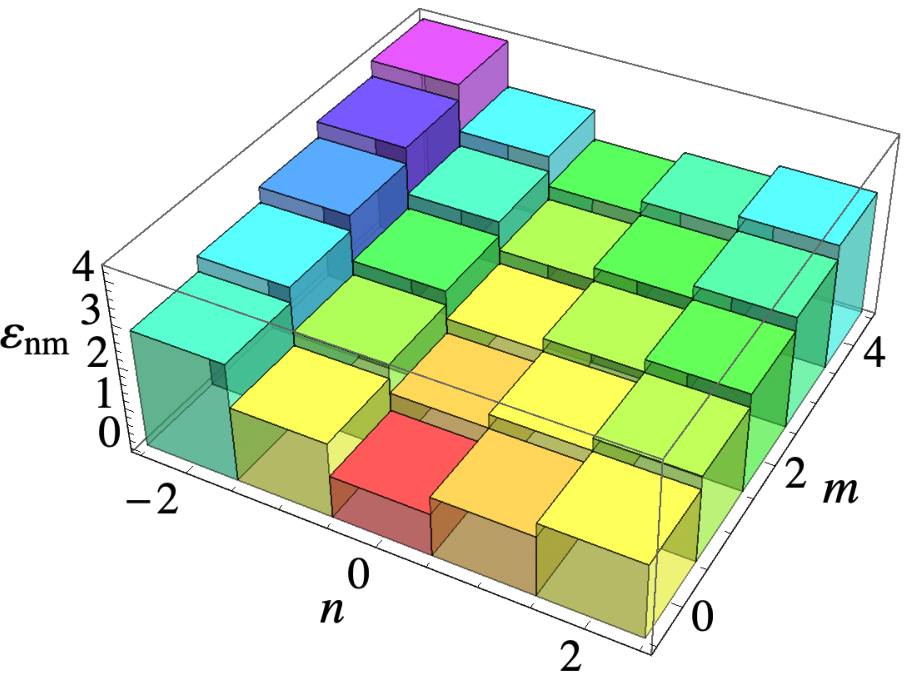}} 
\caption{(a) For square well boundary conditions with $L_2=-3.1\sqrt{2}$ and $L_1=2.1 \sqrt{2}$ (in units of $\ell$) we plot $\vert \mbox{Det}[A_n(u_0)] \vert$ as function of $u_0$ for $n=0$ (blue) and $n=1$ (orange).  The zeroes $u_{nm}$ fix the bound-state energies, $\veps_{nm}=u_{nm} \pm \sqrt{n}, n=0,1,2, \ldots, m=0,1,2,\ldots$.  (b) 3D discrete plot of the bound-state energies $\veps_{nm}$ (negative $n$ means $\veps_{nm}=u_{nm} - \sqrt{n}$).  The points $(n,m,\veps_{nm})$ are the center of a unit square placates with half integer vertices, $(n\pm 1/2,m\pm 1/2)$.  The square placates are drawn simply to graphically clarify the values of $\veps_{nm}$.
%Mathematica file Redpoints-3D.nb 
%{\color{red} Mathematica Files Klein-Shifted-(+x/2)-v0-every-n.nb  and  Klein-u0-Redpoints.nb}
}
\label{Detzero0}
\end{figure}

{\it Matching Equations}.---
Following Eqs.~(\ref{center}) and (\ref{right-left}), for fixed $\pm \sqrt{n}$, the wavefunction is determined by the  coefficients vector ${\bf c}=(c_1,c_2,c_3,c_4)^T$.  Continuity requires
$\psi_{\text{c}}(L_1) = \psi_{\text{right}}(L_1)$ and $\psi_{\text{c}}(L_2) = \psi_{\text{left}}(L_2)$, where each relation yields two equations. This set of four linear  homogeneous equations can be formally written as $A_n(u_0) {\bf c}=0$.  The potential strength $u_0$ must satisfy Det[$A_n(u_0)] = 0$, and the roots $u_{nm}$ determine the bound-state energies $\veps_{nm}=u_{nm} \pm \sqrt{n}$. The eigenvector ${\bf c}_{nm}$ of $A_n(u_{nm})$ corresponding to eigenvalue zero determines the wavefunction in all space.  Figure~\ref{Detzero0}(a) plots Det$[A_n(u_0)]$ versus $u_0$.  For each $0 \le n \in \mathbb{Z}$ there are, in principle, an infinite number of zeros $\{ u_{nm} \}$ and infinite number of bound-state energies $\veps_{nms}=u_{nm}+s \sqrt{n}$,  where $s =\pm$.   A few bound state energies  are shown in Fig.~\ref{Detzero0}(b).

\begin{figure}%[hrb]
\subfigure[]
{\includegraphics[width=0.7\linewidth,angle=0] {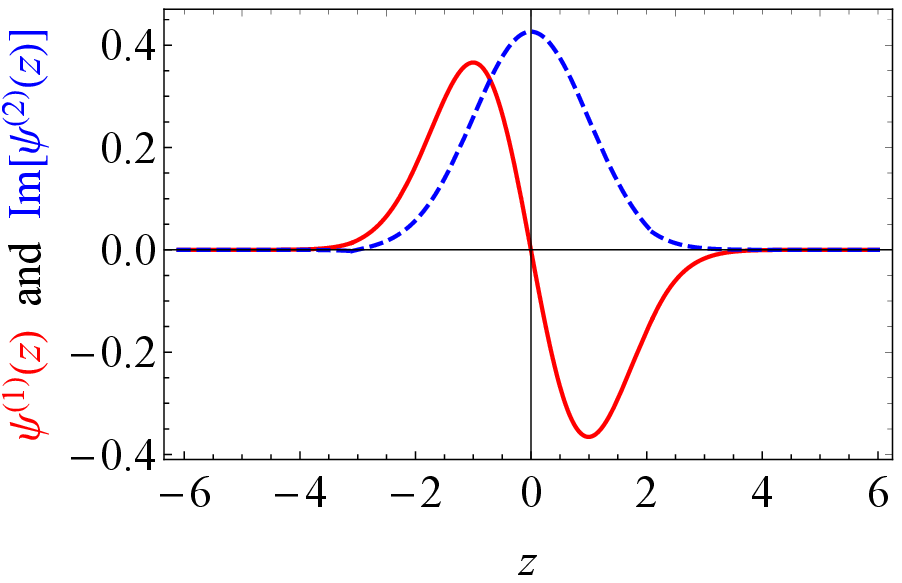}}
\ \ \ \subfigure[]
{\includegraphics[width=0.7\linewidth,angle=0] {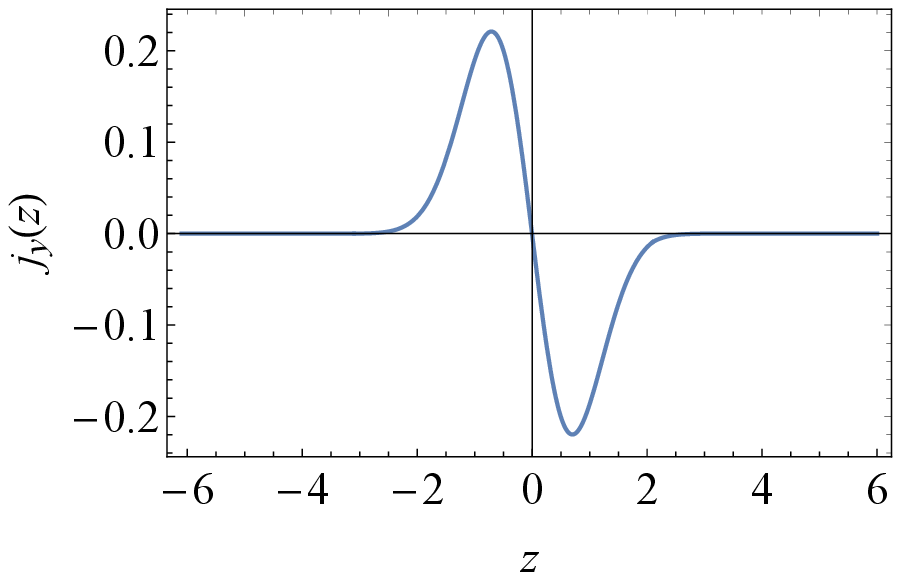}}
%\subfigure[]
%{\includegraphics[width=0.4\linewidth,angle=0]  {psi01.eps}}
%\ \ \ \subfigure[]
%{\includegraphics[width=0.4\linewidth,angle=0]  {cur01.eps}}
\caption{For the potential strength $u_0=u_{n=0,m=0}=1.0013$ [the first blue zero in Fig.~\ref{Detzero0}(a)] and width $L_2=-3.1\sqrt{2}, \ L_1=2.5\sqrt{2}$ we plot (a) the upper component (red) and $-i$ times the lower component (blue) of the wavefunction $\psi_{10}(z)$, and (b) the current $J_y(z)$ of the state $\psi_{00}(z)$. Since $u_{00}$ is small, the wave functions and current 
seem to have perfect symmetry around $z=0$ but, strictly speaking,  they are not (see see discussion in the text and Ref.\cite{SM}). 
%(c) Real and imaginary parts of the wavefunction $\psi_{01}(z)$. (d) Current $J_y(z)$ of the state $\psi_{01}(z)$
%{\color{red} Klein-Shifted-(+x/2)-v0-every-n.nb, Klein-Mag-Derivatives-u.nb, Collected-Results-from-v0-every-n.nb}
}
\label{psi00}
\end{figure}

{\it Wavefunctions and Currents}.---
The spinor wavefunctions and the currents along $y$ corresponding to well height $u_{nm}$
for $(n,m)=(0,0)$  are shown in Fig.~\ref{psi00}. The main properties of the wavefunctions are: (1) It is possible to choose the phase such that the upper component of the spinor is real while the lower component is imaginary. This implies that the current along $x$ vanishes, as it should for bound states. (2) Parity symmetry (or antisymmetry) is not exact for the wave functions around $z=0$. The density $\rho(z)= \psi^\dagger_0(z) \psi_0(z)$  is not perfectly symmetric and the current density $J_y(z)=\psi^\dagger_0(z) \sigma_y \psi_0(z)$ is not perfectly antisymmetric, hence the total (integrated) current $I_y$ does not vanish.
(With the particular choice of parameters adopted here we get $I_y=0.00737326$).  
The reason for this is that the energy levels are degenerate $\veps(k_y)=\veps(-k_y)$ and 
the corresponding quantities for $\pm k_y$ are related:
\begin{equation} \label{rhopsi}
 \rho(z;-k_y)=\rho(-z;k_y), \ \ J_y(z;-k_y)=-J_y(-z;k_y).
\end{equation}
Hence, the (incoherent) weighted sums of contributions from $\pm k_y$ satisfy the pertinent symmetries, and hence $I_y=0$ for the weighted sums. 
In principle, $\rho(z)$ and $J_y(z)$ can be measured 
together with dipole transition rates (see discussion and illustration in the SM \cite{SM}). Therefore, graphene Landau wavefunctions with non-integer orders can be probed.

{\it Summary and Conclusions}.---
We have developed a formalism for studying electron Klein bound states in single layer graphene subject to a symmetric 1D square-well potential, in the absence as well as in the presence of an external magnetic field. This study completes and adds  novel concepts to the analysis of chiral tunneling reported Ref.~\cite{Katsnelson_06}. In the absence of magnetic field, an analytic expression is derived for the wavefunctions of the ground and excited states, and a beautiful symmetry  between the two components of the (pseudo-)spinor is exposed. The consequences of parity non-invariance and time reversal invariance are elucidated, and photon absorption inducing $E1$ transition between two levels is worked out.  In the presence of an external uniform perpendicular magnetic field, an analytic expression for the wavefunctions is derived for a discrete (albeit infinite) sequence of potential strengths  $u_0=\{u_{nm} \}$ $n,m=0,1,2,\ldots$. The Landau functions (in graphene) with non-integer orders and imaginary argument appearing in Eq.~(\ref{center}) are thereby
exposed to experimental probes.  Exact numerical calculations valid for every potential strength are carried out in the supplemental material \cite{SM}, and the importance of the symmetry (\ref{rhopsi}) is stressed. 

Our results apply directly to the propagation of light waves in periodic waveguide optical structures.  Transport of light in a 2D binary photonic superlattice with two interleaved lattices $A$ and $B$ is realized by a sequence of equally spaced waveguides with alternating deep/shallow peak refractive index changes.  Propagation of monochromatic light waves is well-described by the scalar wave equation in the paraxial approximation.  The tight-binding limit results in coupled-mode equations for the fundamental-mode field amplitudes which are functions of a discrete set of integer variables, and approximating these with a continuous variable rather than as an integer index yields a 2D Dirac equation with an external electrostatic potential \cite{Longhi_10, Longhi_11}. This yields the same mathematical formalism used to describe graphene.

{\bf Acknowledgments:} We would like to thank Mikhail I. Katsnelson, Jean N\"oel Fuchs Ken Shiozaki and Ady Stern for illuminating discussions.  This work was supported in part by a grant from the DFG through the DIP program (FO703/2-1).

\end{document}

\title{Supplemental Material for ``Klein Bound States in Single-Layer Graphene''}
\author{Y. Avishai$^{1,3,4}$ and Y. B. Band$^{1,2}$}
\affiliation{$^1$Department of Physics and The Ilse Katz Center for Nano-Science, Ben-Gurion University of the Negev, Beer-Sheva, Israel.\\
  $^2$ Department of Chemistry, Ben-Gurion University of the Negev, Beer-Sheva, Israel.\\
  $^3$ New-York Shanghai University, 1555 Century Avenue, Shanghai, China. \\
  $^4$Yukawa Institute for Theoretical Physics, Kyoto, Japan.\\
}
\maketitle

Here we elaborate on several points discussed in the main text (MT) \cite{main}. Section \ref{sec:matrix} contains additional information regarding the matrix $A(\veps)$ introduced in Eq.~(6) of the MT.  Section \ref{Bne0_numer} considers the numerical solution of Eq.~(18) in the MT, and Sec.~\ref{sec:E1} discusses the electric dipole ($E_1$) transitions between bound states.

\section{The Matrix $A(\veps)$ Related to Eq.~(6)} \label{sec:matrix}

In this section we give an explicit expression for the matrix $A(\veps)$ that determines the bound state energies and wavefunction coefficients specified by the vector ${\bf c} \equiv  (a, b, \alpha, \delta)$ appearing in Eq.~(2).  The pertinent quantities are introduced in the discussion near Eqs.~(3), (4), and (5) in the MT. 
%For self-consistence, some equations are repeated. 
%The (dimensionless) momenta inside the well ($u_0=0$) and outside  the well ($u_0>\veps>0$) are,
%\begin{equation} \label{kF}
% k_x =\veps \cos \phi , \ \  k_y =\veps \sin \phi, \ \ \tan \phi=\frac{k_y}{k_x}, \ \ 
% q_x=\sqrt{(\veps-u_0)^2-k_y^2}, \ \ \tan \theta=\frac{k_y}{q_x}.
% \end{equation}
% In the $p$-$n$-$p$ junction analyzed here, the Klein paradox occurs if the inequality $u_0>\veps >0$ is satisfied. Klein bound-states occur under the more stringent inequality,  
%\begin{equation}  \label{qim}
%u_0>\veps > u_0/(1+\sin \phi)>0 \ \Rightarrow q_x=i \kappa_x=i \sqrt{(\veps \sin \phi)^2 - (u_0-\veps)^2},
%\end{equation}
%where $\kappa_x >0$ (real and positive).
%The bound-state wave functions must decay exponentially as $e^{-\kappa_x \vert x \vert}$ as  $\vert x \vert \to \infty$. In this region $\tan \theta= -i k_y/\kappa_x$ is pure imaginary, and $\tan^2 \theta <-1$. 
%To insure this behaviour we must set $\beta=\gamma=0$ in Eq.~(\ref{psi-bound-state-x}), keeping only the decaying parts of the wave function for $\vert x \vert >1$. 
The matching conditions at $x=\pm 1$ lead to a homogeneous  system of four linear equations for the complex coefficients $a, b, \alpha, \delta$.  Bound-state solutions occur at energies $\{ \veps_n \}$ for which the determinant of $A(\veps)$ vanishes, and the corresponding coefficient vector ${\bf c}_n$ is determined by the set of homogeneous equations $A(\veps_n){\bf c}_n=0$. The explicit form of the matrix $A(\veps)$ in the system of equations, $A(\veps) {\bf c}=0$, is given by
%{\color {red} Klein-Bound-State-Matrix-Eigensystem2.nb}
\begin{eqnarray} \label{MatrixA}
 && A(\veps)=  
  \left( \begin{array}{cccc}
 e^{i  \veps \cos \phi} & e^{-i  \veps \cos \phi} & -e^{-\kappa_x} & 0 \\
 e^{i \left( \veps \cos \phi+\phi \right)} & -e^{-i \left( \veps \cos \phi+\phi
   \right)} & e^{ \left( -\kappa_x+i\theta \right)} & 0 \\
 e^{-i  \veps \cos \phi} & e^{i  \veps \cos \phi} & 0 & -e^{-\kappa_x} \\
 e^{-i \left( \veps \cos \phi-\phi \right)} & -e^{i \left( \veps \cos \phi-\phi
   \right)} & 0 & -e^{ \left(- \kappa_x-i \theta \right)} \\
\end{array} \right).
\end{eqnarray}

%\section{$B \ne 0$: Potential strengths for which Eq.~(\ref{right-left}) is satisfied.} 
%\label{sec:Bne0}

\section{Numerical solution of Eq.~(18)}  \label{Bne0_numer}

The set of Landau functions is complete on the interval $(-\infty,\infty)$ so we can expand $\psi(z)$:
\begin{eqnarray} \label{EF1}
&& \psi(z)=\sum_{n=0}^{M \to \infty} \sum_{s=\mp} a_{ns} L_{ns}(z), 
%\nonumber \\ 
%&& 
~ ~ \mbox{where} \ \ L_{ns}(z)=N_{ns} \begin{pmatrix} \! D_{n}(z), \  \\ 
s i \sqrt{n} D_{n-1}(z) \end{pmatrix},
\end{eqnarray}
Here $N_{ns}$ is a normalization factor. Substitution into Eq.~(18) then yields, 
\begin{eqnarray} \label{EF2}
 && [-i \sigma_x \partial_z -\tfrac{1}{2}z\sigma_y]\psi(z)=\sum_{n=0}^M \sum_{s=\mp} a_{ns} s \sqrt{n} L_{ns}(z) 
 %\nonumber \\
 %&&
 =[\veps-u(z)]  \sum_{n=0}^M \sum_{s=\mp} a_{ns} L_{ns}(z).
\end{eqnarray}
Multiplying by $L_{m t}^\dagger(z)$ (where $t=\mp$) and integrating over $z$,
using $\la L_{mt}\vert L_{ns} \ra=\delta_{mn} \delta_{ts}$ one obtains,\\
%{\color{red} REWRITE Eq.~(\ref{EF3}) so it looks like an eigenvalue problem}
\begin{equation} \label{EF3}
  t \sqrt{m}a_{mt}=\veps a_{mt}-\sum_{n=0}^M\sum_{s=\pm} A_{mt,ns}a_{ns}.
\end{equation}
The infinite sum (as $M \to \infty$) can be cut-off at a sufficiently large $M$. This procedure leads to an eigenvalue problem in a finite Hilbert space of dimension $2M+1$. The matrix $A$ introduced above can be written as $u_0(I-B)$, where $I$ is the $(2M+1)$$\times$$(2M+1)$ unit matrix. The explicit expressions for the matrices $A$ and $B$ are,
\begin{eqnarray} \label{EF4}
&& A_{mt,ns}=\int_{-\infty}^\infty L_{m t}^\dagger(z)u(z)L_{ns}(z) dz = u_0[\underbrace{ \delta_{mn} \delta_{ts}}_{I_{mt,ns}} - 
\underbrace{\int_{L_2}^{L_1} L^\dagger_{mt}(z) L_{ns}(z) dz}_{B_{mt,ns}} ].
\end{eqnarray}
where $u_0$ is the strength of the square well potential defined in Eq.~(17) in the MT.  Next, we define a diagonal matrix $\Lambda$ by 
\begin{eqnarray} \label{EF5}
  &&\Lambda_{mt,ns}=\delta_{mt,ns}\mbox{Diag}(t \sqrt{m}) 
  %\nonumber \\
  %&&
  =(0,\sqrt{1},\sqrt{2},\ldots,\sqrt{M},-\sqrt{1},-\sqrt{2}, \ldots,-\sqrt{M}),
\end{eqnarray}
and a vector ${\bf a}$ with $2M+1$ components,
%\begin{equation} \label{EF6} 
$a_{ns}=(a_{0},a_{1+},a_{2+},\ldots,a_{M+},a_{1-},a_{2-}\ldots,a_{M-})$.
%\end{equation}    
Equation~(\ref{EF3}) then becomes an eigenvalue problem,
\begin{equation} \label{EF7}
  [\Lambda+u_0(I-B)]{\bf a}=\veps {\bf a},
\end{equation}
The matrix $\Lambda+u_0(I-B)$ is real and symmetric. For $u_0=0$ the eigenvalues are the Landau energies for graphene $\veps_m=\pm \sqrt{m}$.  In the calculations of density, current and $E_1$ transitions presented below we take $u_0=\frac{\ell U_0}{\hbar v_F}=10$, $-\frac{5}{2 \sqrt{2}} \le x \le \frac{5}{2\sqrt{2} } $ (in units of $\ell$),  and $k_y=\pm 0.5$ (in units of $1/\ell$). Since $z=\sqrt{2}x-k_y$, this gives $[L_2,L_1]=[-3,2]$ for $k_y=+0.5$, and $[L_2,L_1]=[-2,3]$ for $k_y=-0.5$.

\begin{figure}%
\centering
\subfigure[ ]{%
\includegraphics[height=1.7 in]{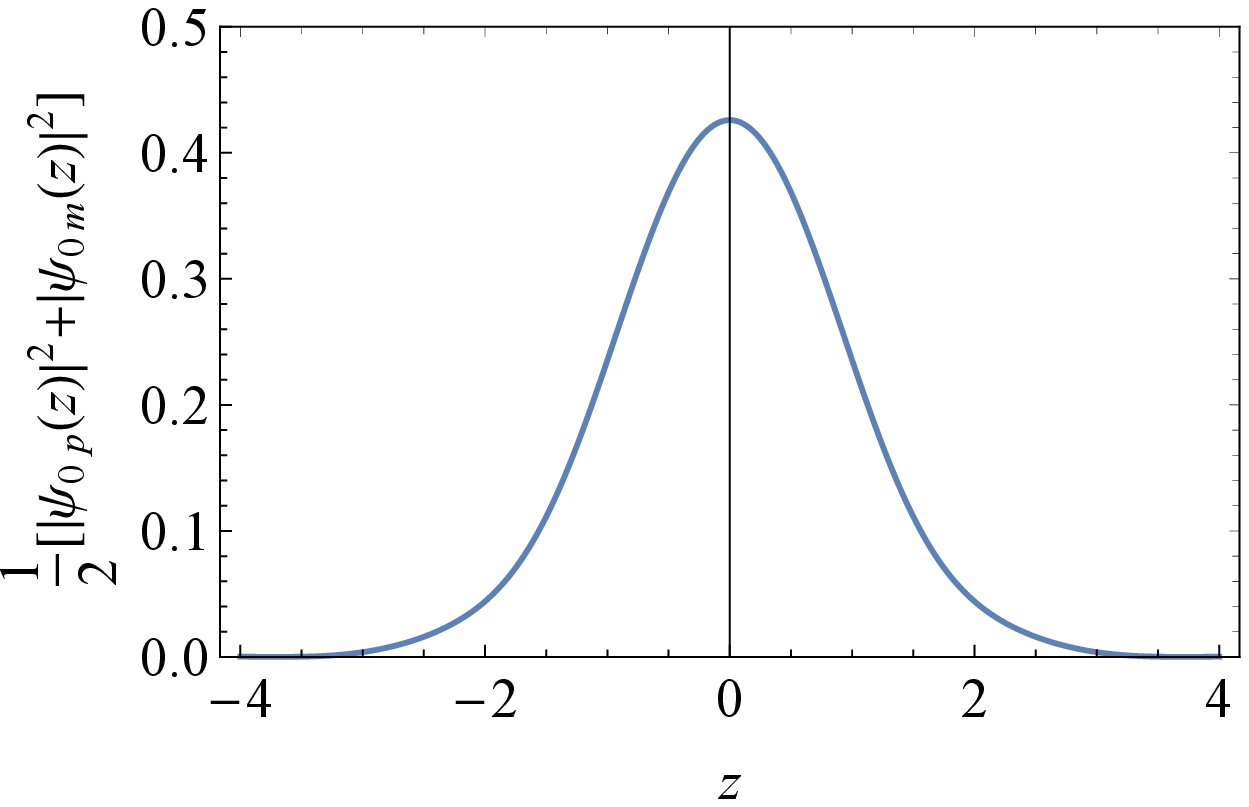}}%
\subfigure[]{%
\ \ \ \includegraphics[height=1.7 in]{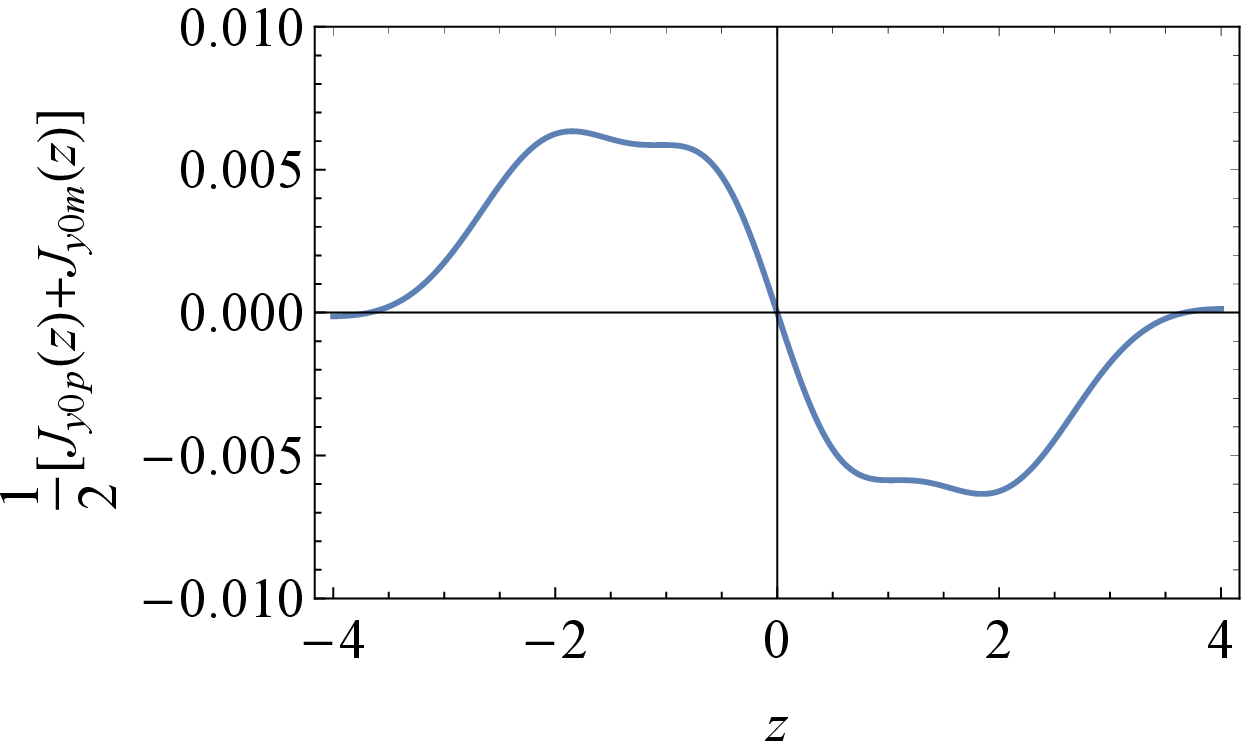}}%
\caption{ Density $\rho_0(z)$ and current $J_{y0}(z)$ 
of  the ground state $\psi_0(z)$ following incoherent summation over $\pm k_y$. 
(a) $\tfrac{1}{2}[\rho_0(z,k_y )+\rho_0(z,-k_y )$, (b)$\tfrac{1}{2}[J_{y0}(z,k_y )+
J_{y0}(z,-k_y )$ .}
\label{psi0}
\end{figure}

\begin{figure}%
\centering
\subfigure[ ]{%
\includegraphics[height=1.7 in]{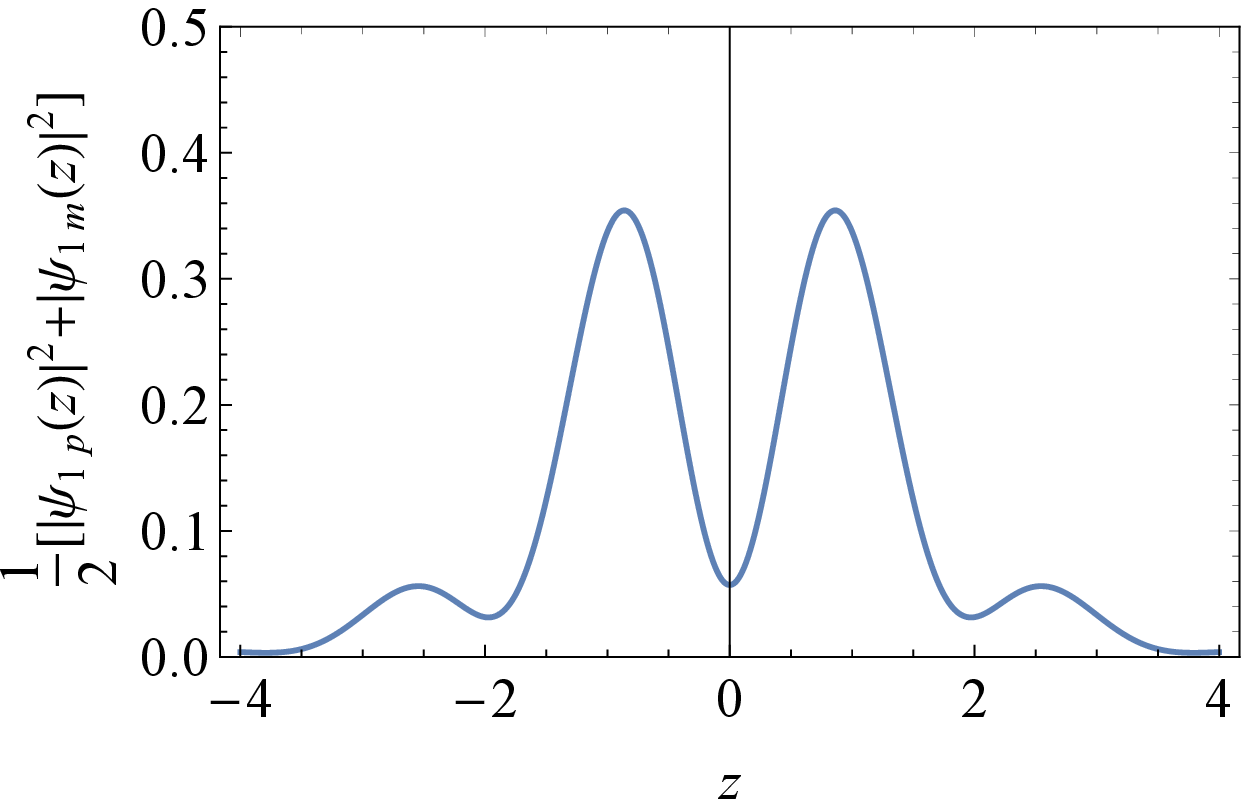}}%
\subfigure[]{%
\ \ \ \includegraphics[height=1.7 in]{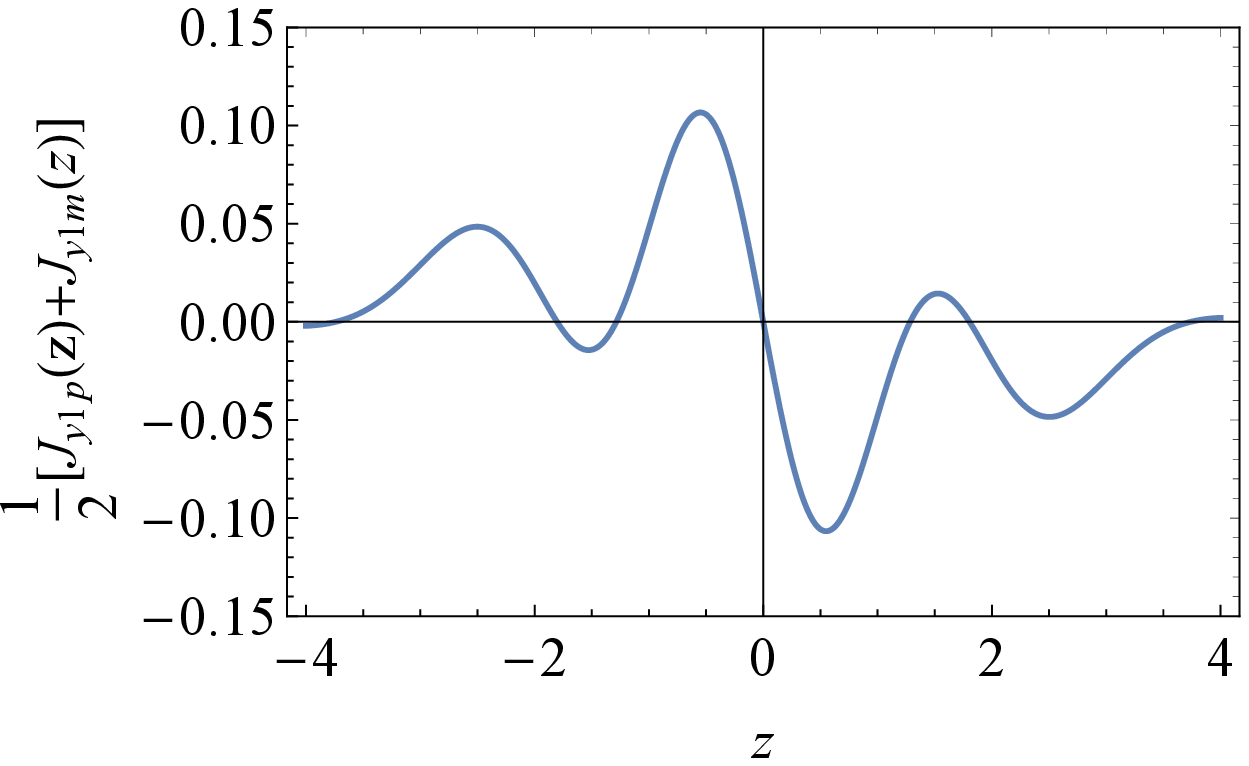}}%
\caption{ Density $\rho_1(z)$ and current $J_{y1}(z)$ 
of the first excited state $\psi_1(z)$ following incoherent summation over $\pm k_y$. 
(a) $\tfrac{1}{2}[\rho_1(z,k_y )+\rho_1(z,-k_y )$, (b)$\tfrac{1}{2}[J_{y1}(z,k_y )+
J_{y1}(z,-k_y )$ .}
\label{psi1} 
\end{figure}

Figure \ref{psi0} shows the ground-state density symmetrized density $\rho_0(z)=\tfrac{1}{2} \sum_{\pm k_y}[\psi_0^\dagger(z) \psi_0(z)]$  and current density along $y$, 
$J_{y0}(z)=\tfrac{1}{2} \sum_{\pm k_y}[\psi_0^\dagger(z) \sigma_y \psi_0(z)]$. As argued in the discussion of Eq.~(21) of the MT, the incoherent sum of contributions from $\pm k_y$ results in a symmetric density and an antisymmetric current density. In particular, the total current along $y$, $I_{y0}=\int_{-\infty}^\infty J_{y0}(z) dz$ vanishes (as it should). Similar results for the first excited state $\psi_1(x)$ are shown in Fig.~\ref{psi1}. 
%%%%%%%%%%%%%%%%%%%%%%%%%%%%%

\section{$E_1$ transitions in the presence of magnetic field}  \label{sec:E1}

\begin{figure}[htb]
\includegraphics[width=0.4\linewidth,angle=0] {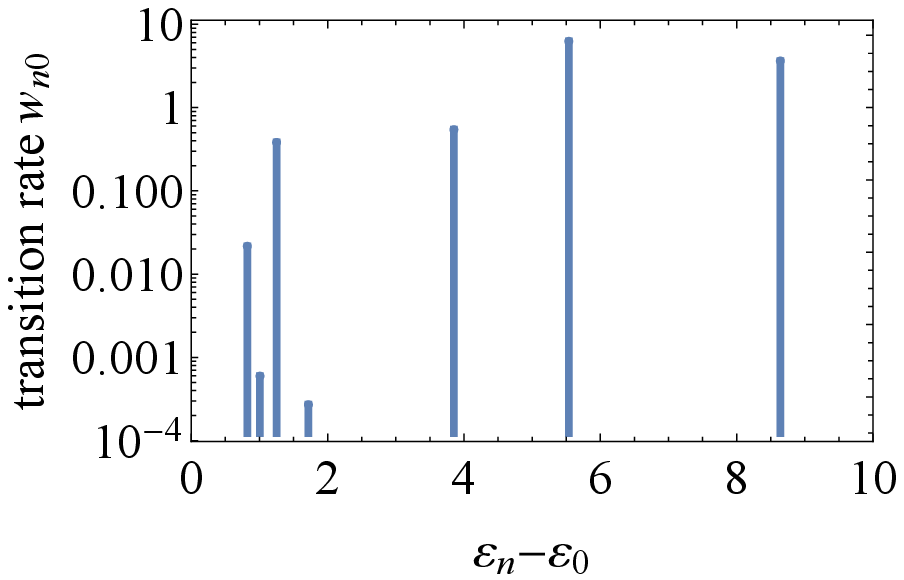}
\caption{Absorption spectrum of the transitions $0 \to n$, $n=1,2,\ldots,7$.  The transition rates $w_{0,n}$ (in dimensionless units), which are proportional to $\omega_{0n}^4 \, \left| \langle \psi_0| x |\psi_n \rangle \right|^2$, are plotted versus the resonant light photon energy $\hbar \omega_{0n}=\veps_n-\veps_0$ in dimensionless units.  
%{\color{red} See text for transforming dimensionless quantities to physical ones.}
}
\label{absorption_spectrum-B}
%{Klein-Modified-Determinant-band4-ky-new.nb}
\end{figure}

In analogy with the discussion of photon absorption in the absence of an external magnetic field [see Eq.~(14) in the MT], we now consider $E_1$ transitions in the presence of the magnetic field. The $E_1$ transition rates $w_{n,m}$ from $m$ to $n$ with light polarized along the $x$ axis are proportional to $|\veps_n-\veps_m|^4  \, \left| \langle \psi_n| x |\psi_m \rangle \right|^2$, where $\{ \veps_n \}$ are the energy eigenvalues obtained from the solution of Eq.~(\ref{EF7}) and the transition dipole matrix elements are
\begin{equation}   \label{EF8}
\la x \ra_{mn} = \la \psi_{m}\vert x \vert \psi_{n} \ra = \int_{-\infty}^\infty\psi_{m}^\dagger[z(x)] x \psi_n[z(x)]dx,
\end{equation} 
where $z(x)=\sqrt{2}(x-k_y)$. The main contribution comes from the interval $-L \le x \le L$ where $L/\ell=\frac{5}{2 \sqrt{2}}$ [see details below Eq.~(\ref{EF7})].  Photon absorption spectrum between the lowest eight states $n=0,1,\ldots,7$ [determined  within the set of parameters specified after Eq.~(\ref{EF7})], is shown in Fig.~\ref{absorption_spectrum-B}. 
It is interesting to underline the differences between photon absorption spectra in the 
presence and in the absence of the magnetic field shown in Fig.~2 in the MT. In the latter case, there is the usual parity selection rule, namely, the function $\psi^\dagger_n(x) x \psi_m(x)$ is even (odd) if $n+m+1$ is odd (even). In particular, transitions $0 \to 1,3,5,7$ are shown but $0 \to 2,4,6$ vanish. These parity selection rules do not apply in the presence of magnetic field, hence all transitions $0 \to n$ ($n=1,2,\ldots,7$) are allowed. 
  
%{\color {red} Klein-B-Eigenfunction-YB-pm-ky-E1}.

One can easily convert the $E_1$ transition rates to physical units.
Following Ref.~\cite{BAQMN} (p.~324), the $E_1$ transition rate between states $|m \ra$ and $|n \ra$ is  given, (up to a multiplicative factor $A$ depending on constants such as $c$ and $\hbar$), by:
$$w_{mn}=A (E_m-E_n)^4 (e {\cal E}_x)^2 \vert X_{mn}\vert^2, \ \ X_{mn} = \la m \vert x \vert n \ra,$$
where $E_m$ and $E_n$ are the level energies and ${\cal E}_x$ is the slowly varying envelope of the electric field. The physical dimension of $w_{mn}$ is $[w_{mn}]=[A] \times \mbox{[energy]}^6$. If there is a parameter of length $d$ in the system, then we can use it as the unit of length and work with dimensionless quantities:
$\veps_m$ for energy $E_m$,  $u_0$ for potential height $U_0$, $k_y \to k_y d$ for wave numbers and $x_{mn}$ for $X_{mn}$: 
$$ E_m=\frac{\hbar v_F}{d} \veps_m \ \  , \ U_0=\frac{\hbar v_F}{d}u_0, \  X_{mn}=d x_{mn}, \ \ \Rightarrow w_{mn}=A\left [\frac{\hbar v_F}{L} \right ]^4 (\veps_n -\veps_m)^4 (e {\cal E}_x d)^2 \vert x_{mn}\vert^2. $$
In order to compute $w_{mn}$ we still need to know ${\cal E}_x$ and compute the energy $e {\cal E}_x d$ for these values of $d$ and ${\cal E}_x$. In the absence of an external magnetic field, we can use $d=L$, where $2L$ is the width of the square well. In the presence of a magnetic field, we can use $d=\ell$, the magnetic length. In this case, the dimensionless width of the square well is $2 L/\ell$.  The relevant energies can be inferred by noting that for a magnetic field of 1 T, $\ell \approx 25$ nm and $\hbar v_F/\ell \approx 21.875$ meV.